\documentclass[12pt]{article}         
\usepackage{graphicx}                 
\usepackage{adjustbox}
\begin{document}

\font\fiverm=cmr5                     


\catcode`@=11 \catcode`!=11

\expandafter\ifx\csname fiverm\endcsname\relax
  \let\fiverm\fivrm
\fi
  
\let\!latexendpicture=\endpicture 
\let\!latexframe=\frame
\let\!latexlinethickness=\linethickness
\let\!latexmultiput=\multiput
\let\!latexput=\put
 
\def\@picture(#1,#2)(#3,#4){%
  \@picht #2\unitlength
  \setbox\@picbox\hbox to #1\unitlength\bgroup 
  \let\endpicture=\!latexendpicture
  \let\frame=\!latexframe
  \let\linethickness=\!latexlinethickness
  \let\multiput=\!latexmultiput
  \let\put=\!latexput
  \hskip -#3\unitlength \lower #4\unitlength \hbox\bgroup}

\catcode`@=12 \catcode`!=12
\input{pictex.tex}

 
\catcode`@=11 \catcode`!=11
  
\let\!pictexendpicture=\endpicture 
\let\!pictexframe=\frame
\let\!pictexlinethickness=\linethickness
\let\!pictexmultiput=\multiput
\let\!pictexput=\put

\def\beginpicture{%
  \setbox\!picbox=\hbox\bgroup%
  \let\endpicture=\!pictexendpicture
  \let\frame=\!pictexframe
  \let\linethickness=\!pictexlinethickness
  \let\multiput=\!pictexmultiput
  \let\put=\!pictexput
  \let\input=\@@input   
  \!xleft=\maxdimen  
  \!xright=-\maxdimen
  \!ybot=\maxdimen
  \!ytop=-\maxdimen}

\let\frame=\!latexframe

\let\pictexframe=\!pictexframe

\let\linethickness=\!latexlinethickness
\let\pictexlinethickness=\!pictexlinethickness

\let\\=\@normalcr
\catcode`@=12 \catcode`!=12



\newcounter{bibno}

\def\waux{1}

\newcommand\writelabel[2]
   {
     \immediate\write\waux
          {\noexpand\newlabel{#1}
             {{#2}{\thepage}}}
   }

\newcommand\bibref[1]
   {%
    \addtocounter{bibno}{1}
    \def\biblab{\thebibno}
    \writelabel{#1}{\biblab}
    \noindent\biblab.
   }%


\newcommand{\0}{\mbox{\boldmath $0$}}         
\newcommand{\bbeta}{\mbox{\boldmath $\beta$}} 
\newcommand{\bd}{\mbox{\boldmath $d$}}        
\newcommand{\bg}{\mbox{\boldmath $\gamma$}}   
\newcommand{\blam}{\mbox{\boldmath $\lambda$}}
\newcommand{\bpi}{\mbox{\boldmath $\pi$}}     
\newcommand{\bsig}{\mbox{\boldmath $\sigma$}} 
\newcommand{\bth}{\mbox{\boldmath$\theta$}}   
\newcommand{\dt}{\widetilde{d}}               
\newcommand{\m}{\mbox{\boldmath $m$}}         
\newcommand{\M}{\mbox{\boldmath $M$}}         
\newcommand{\mi}{{$-$}}                       
\newcommand{\n}{\mbox{\boldmath $n$}}         
\newcommand{\ndot}{n{\rm\bf .}}               
\newcommand{\Nh}{\widehat{N}}                 
\newcommand{\nsp}{\hspace*{-.3in}}            
\newcommand{\p}{\phantom{5}}                  
\newcommand{\pih}{\widehat{\pi}}              
\newcommand{\pmi}{\phantom{$-$}}              
\newcommand{\ps}{\phantom{$*$}}               
\newcommand{\pc}{{\scriptsize$\bullet$}}      
\newcommand{\q}{\phantom{55}}                 
\newcommand{\qs}{\hspace*{.1in}}              
\newcommand{\re}{{\rm e}}                     
\newcommand{\bt}{\mbox{\boldmath $t$}}        
\newcommand{\bT}{\mbox{\boldmath $T$}}        
\newcommand{\U}{\mbox{\boldmath $u$}}         
\newcommand{\X}{\mbox{\boldmath $X$}}         
\newcommand{\x}{\mbox{\boldmath $x$}}         
\newcommand{\Y}{\mbox{\boldmath $Y$}}         
\newcommand{\y}{\mbox{\boldmath $y$}}         
\newcommand{\z}{\mbox{\boldmath $z$}}         


\hyphenation{general-iza-tions}
\hyphenation{distri-bution}
\hyphenation{Zel-ter-man}


\thispagestyle{empty}
\begin{center}
\vspace*{\fill}

{\Large\bf A Two-Stage, Phase II Clinical Trial Design with} \\
{\Large\bf Nested Criteria for Early Stopping and Efficacy}\\ [1ex]

\vspace*{1.25in}
\begin{tabular}{l} 
 {\bf Daniel Zelterman}  \\[.1in]

 Department of Biostatistics \\
 School of Epidemiology  \\
 \hspace*{.15in} and Public Health  \\
 Yale University       \\
 New Haven, CT  06520  \\[.2in]

\today
\end{tabular}
\end{center}

\vspace*{\fill}

\noindent${}^*$Email address for author: {\tt
daniel.zelterman@yale.edu}.  This research was supported by grants 
	R01CA131301, R01CA157749, R01CA148996, R01CA168733 awarded by the National Cancer Institute, and support from the Yale Comprehensive Cancer Center.
The author thanks Elizabeth Nichols for her editorial support.


\newpage
\thispagestyle{empty}

\section*    {\bf   Abstract}

We propose a two-stage design for a clinical trial with an early stopping rule for safety.  
We use different criteria to assess early stopping and efficacy.
The early stopping rule is based on a criteria that can be determined more quickly than that of efficacy.
These separate criteria are also nested in the sense that efficacy is a special case of, but not identical to, the early stopping criteria.
The design readily allows for planning in terms of statistical significance, power, and expected sample size necessary to assess an early stopping rule.
This method is illustrated with a Phase~II design comparing patients treated  for lung cancer with a novel drug combination to those treated using historical control.  
In this example, the early stopping rule is based on the numbers of patients who exhibit progression-free survival (PFS) at 2~months post treatment follow-up and efficacy is judged by the number of patients who have PFS at 6~months.

\bigskip

\noindent{\bf Keywords:} lung cancer; power; sample size; early stopping rule; discrete distribution


\thispagestyle{empty}
\setcounter{page}{1}

\section                       {Introduction}

A fundamental design feature of the two-stage Phase II clinical trial is the convenient early stopping rule that prevents us from exposing excessive numbers of patients to a possibly inferior treatment. 
In the traditional two stage designs~[\ref{Simon 1989}], we use the same criteria for early stopping as with the final test of efficacy.
Simply, a number of patients are initially enrolled in Stage~1.
Then we wait to see their outcome before enrolling the remainder of the subjects in Stage~2.  
The criteria for advancing from Stage~1 to Stage~2 is based on a (statistically) large number of patients achieving the same favorable criteria in Stage~1 as will be latter examined for efficacy of treatment at the conclusion of Stage~2. 
An ideal choice of criteria of outcome for such a design is one that can be assessed  quickly.  
A good example of favorable criteria might include response to therapy, as determined a few weeks after treatment, verified by a CT scan.  
The motivation for our present work is the setting where the efficacy determination can only be made after a lengthy follow-up time on each patient.

Our motivating problem came about in a study of a novel drug combination treatment of lung cancer in which our efficacy criteria is 6~month PFS in geriatric patients.  
This criteria would create a long delay between Stage~1 and Stage~2 as we wait to observe the outcomes of the Stage~1 patients before enrolling additional patients into Stage~2. 
Using traditional designs~[\ref{Simon 1989}], there could be up to an 6~month delay in accrual until the outcomes of the Stage~1 patients become available.
One solution to this delay would be to enroll all patients without an early stopping rule, but such a design eliminates any safeguards of exposing excessive numbers of patients to an inferior or harmful treatment.
Another solution would be to continue accruing patients into Stage~2 while waiting for those in Stage~1 to achieve their endpoints. 
This approach also has the potential of exposing too many patients to a possibly inferior treatment.

Our proposed solution is to use a criteria for early stopping that is different from the criteria used to assess efficacy. 
We want these separate criteria to be nested in the sense that the efficacy criteria is a special case of the safety criteria.  
That is, the criteria for efficacy can not be achieved if the patient fails the shorter term, safety outcome. 
In our specific application in Section~\ref{guidance.section}, the efficacy measure is 6~month PFS and we will use the rate of 2~month PFS as the criteria for early stopping.
The estimates of 2- and 6-month historical PFS rates are based on a published Kaplan-Meier survival curve $\,\widehat{S}(t)\,$ estimating PFS times.
The published survival curve $\,\widehat{S}\,$ provides a standard-of-care comparison and provides us with the flexibility to choose any pair of PFS times $\,t_1 \leq t_2\,$ and the corresponding survival rates.

The large literature written on two-stage designs attests to their popularity in clinical trial practice.
We cite only some of the literature here. 
The original idea of the two-stage design for Phase II trials is attributed to R.Simon~[\ref{Simon 1989}], with other developments building on the original idea [\ref{Thall 1990}-- 
\ref{Jung 2001}]. 
There are review articles [\ref{Gehan 1990}--
\ref{DeMets 2012}] covering Phase II designs and issues in clinical trials, more generally. 
The use of historical control data suggests Bayesian approaches [\ref{Lee 2012}--
\ref{Banerjee 2006}].  
Estimation of parameters [\ref{Kunz 2012}, \ref{Bowden 2012}] is not a simple task because of the complex sampling distributions that are involved.  
There are generalization to three or more stages~[\ref{Zhong 2012}, \ref{Ensign 1994}].  
There is also some work~[\ref{Hong 2012}, \ref{Chang 2007}] in how Phase~II endpoints can help in planning subsequent Phase~III studies.

The present manuscript is concerned with separate criteria for safety and efficacy.
Thall and Cheng~[\ref{Thall 1999}, \ref{Thall 2001}]  also consider clinical trial designs with separate parameters for safety and efficacy for events that are not nested, as developed here.  
Similarly, small sample size studies with multiple safety observations on each subject are considered as well~[\ref{Shih 2004}].

In Section~\ref{notation.section} we provide the necessary notation and derive all of the mathematical formul\ae.
In Section~\ref{guidance.section} we provide practical guidance using the parameters of the proposed lung cancer trial as an example.
Section~\ref{discussion.section} compares the traditional two-stage design due to Simon [\ref{Simon 1989}] with the designs proposed here.

\section   {Notation and basic results}
\label{notation.section}

We begin the development of the proposed design by considering rates of overall survival or PFS at two follow-up times $\,t_1 \leq t_2$.  
Survival or PFS at two different times is the most obvious example of a pair of nested criteria that could fit the framework of this design.  
Our results are not specific to these choices however.
Nevertheless, we will continue the narrative describing the outcomes as PFS at two different time points.

In typical planning for the Phase II studies proposed here, there will be a published Kaplan-Meier survival (or PFS) estimate $\,\widehat{S}(t)\,$ that we will use as a comparison to the established standard of care.
The new treatment will be compared to this estimate of a historical control. 
The estimated rates of survival, $\,\widehat{p}_i=\widehat{S}(t_i),\; (i=1,2)\,$ obtained from this survival curve provide us with a pair of parameters needed to design our study.
We will omit the carets on $\,\widehat{p}_1\,$ and $\,\widehat{p}_2\,$ but the reader should keep in mind that these historical control rates are usually published  estimates, as is typically the case with Phase II studies.

The null hypothesis that we wish to test is that overall survival or PFS at time $\,t_2\,$ is equal to $\,p_2$ and the alternative hypothesis is that this rate is greater than $\,p_2$.
The survival rate $\,p_1\,$ at time $\,t_1\,$ is a safety parameter that is used to judge the early stopping rule. 
These rates satisfy $\,0\leq p_2 \leq p_1\leq 1.$
In the traditional Simon two-stage design~[\ref{Simon 1989}] we have $\,t_1=t_2\,$ and  $\,p_1=p_2\,$ because the safety and efficacy criteria are identical.
We will discuss these parameters again and provide a specific example in Section~\ref{guidance.section}.

We will need to identify study parameters for a design that terminates the clinical trial early if our observed survival fraction at time $\,t_1\,$ is not at least as large as $\,p_1.$  
Let $\,n_1\,$ denote the sample size of patients enrolled in Stage~1 of the study.  
Let the random variable $\,X_1\,$ denote the number of these patients who have successful outcomes after each has been followed for a duration of time $\,t_1$.

The random variable $\,X_1\,$ has a binomial distribution with parameters $\,n_1\,$ and $\,p_1$.  
For an {\it a priori\/} fixed cut off value $\,r_1,$ if $\,X_1\geq r_1\,$ then the trial will continue from Stage~1 into Stage~2.
The design parameter $\,r_1\,$ will need to be determined as part of the clinical trial planning process.   
The $\,X_1\,$ successful patients in Stage~1 will continue to be followed up to time $\,t_2\,$ if we decide to proceed into Stage~2.
We will come back to these subjects in a moment.
If $\,X_1 < r_1\,$ then the trial will terminate early.

In Stage~2 we enroll additional patients, follow each of them for a longer period of time $\,t_2$, and use their responses to test the null hypothesis for efficacy.
Specifically, in Stage~2, an additional $\,n_2\,$ patients will be enrolled and each will be followed for a minimum period of time $\,t_2\,$ where $\,t_2\geq t_1$.  
Let the random variable $\,X_2\,$ denote the number of successes among these Stage~2 patients. 
The distribution of $\,X_2\,$ is binomial with index parameter $\,n_2\,$ and probability of success $\,p_2,$ again assuming the historical rates for comparison as the null hypothesis.

In addition to these $\,X_2\,$ successes, there will be a random variable, denoted by $\,X_{12},$ of successes at time $\,t_2\,$ who were initially enrolled in Stage~1.
That is, $\,X_{12}\,$ is the conditional number of successes at time $\,t_2\,$ given that they were successes at time $\,t_1$.
At the conclusion of Stage~2, we will declare the new treatment successful (reject the null hypothesis) if the total number of successes $\,(X_{12} + X_2)\,$ at time $\,t_2\,$ exceeds a study parameter $\,r_2\,$ that is determined {\it a priori}.
Fig.~\ref{schema.fig} is a useful representation of the time line for this design.

Given these notations, we can proceed with the necessary derivations for evaluating the operating characteristics of a proposed clinical trial design of this form.

The probability of early termination is 
$$
    \Pr[\,{\rm Early \ termination }\,]=
     \Pr[\, X_1 < r_1\, ]
$$
where $\,X_1\,$ has a binomial distribution with index $\,n_1\,$ and probability parameter $\,p_1$.

From this expression we also have a bound on the expected sample size:
\begin{equation}                                      
  {\rm Expected \ sample \ size} \leq n_1 +     \label{ess.eq}
          n_2\,(1-\Pr[\,{\rm Early \ termination}\,]).
\end{equation}

Expression~(\ref{ess.eq}) is an upper bound on the expected sample size.
A more useful estimate models the rate of accrual necessary until $\,r_2\,$ successes are observed at follow-up time $\,t_2$.
In the Appendix, we provide details for obtaining the expectation of the minimum number of Stage~1 patients needed in order to be able to make a decision whether or not to continue into Stage~2.

Let us next derive an expression for the probability that we reject the null hypothesis (H${}_0$):
$$
 \Pr[\, {\rm reject \, H}_0\, ] = \Pr[\,X_{12} + X_2 \geq r_2\,] .
$$

We first condition on the number of successes $\,X_1\,$ in Stage~1:
$$
 \Pr[\, {\rm reject \, H}_0\,] =
          \sum_{x_1} \;\Pr[\, X_1=x_1\,]\,
               \Pr[\, X_{12}+X_2 \geq r_2\mid  x_1\,] 
$$

Since $X_{12}\,$ and $\,X_2\,$ are independent, we can write the distribution of their sum as a convolution:
$$
   \Pr[\,{\rm reject \, H}_0\,] = 
       \sum_{x_1} \;\Pr[\,X_1=x_1\,]\,
       \sum_{x_{12}}\, \Pr[\,X_{12} = x_{12}\mid x_1\,]
              \;\;    \Pr[\,X_2 \geq r_2-x_{12}\,]
$$

The conditional distribution of $\,X_{12}\,$ given $\,X_1\,$ is binomial with index parameter $\,X_1\,$ and probability parameter equal to $\,p_2/p_1$. 
The distribution of $\,X_2\,$ is independent of both $\,X_1\,$ and $\,X_{12}$.

This gives us
\begin{eqnarray}                                 
\hspace*{-.35in}\Pr[\, {\rm reject \, H}_0\, ] 
       &=& \nonumber\\[1ex] && \hspace*{-.95in}
         \sum_{\shortstack{\\[0ex] $x_1=$ \\[.5ex]
                      $\max(r_1,r_2-n_2)$}}
                   ^{\shortstack{$n_1$}} 
       \hspace*{-.25in} \Pr[\,X_1=x_1\,]
       \hspace*{-.2in}\sum_{\shortstack{ \\ $x_{12}=$ \\
                      $\max(0,r_2-n_2)$}}
                   ^{\shortstack{$x_1$}}
       \hspace*{-.2in}\Pr[\,X_{12}=x_{12}\mid x_1\,]
       \hspace*{-.25in}\sum_{\shortstack{\\ $x_2=$ \\ 
                    $\max(0,r_2-x_{12})$}}
                   ^{\shortstack{$n_2$}}
       \hspace*{-.25in}\Pr[\, X_2=x_2\,]   \label{pr.reject.eq}
\end{eqnarray}

The limits of the summation must take into account both the decision to continue sampling after Stage~1 (i.e. $\,X_1 \geq r_1$) as well as the rejection of the null hypothesis at the end of Stage~2 (i.e. $\,X_{12}+X_2\geq r_2$).
The lower limits of each summation in~(\ref{pr.reject.eq}) are not greater than the upper limits so long as there are valid design criteria in $\,n_1, n_2, r_1, {\rm \ and \ } r_2$.  
The limits and ranges for these parameters are summarized in Table~\ref{parameters.table}. 
This table includes a summary of the distributions needed to compute this probability.
The {\bf R} programs are also available by request from the author.

For specified values of $\,(p_1, p_2)\,$ we find values of the design parameters $\,(n_1,n_2,r_1, r_2)\,$ so that the probability in~(\ref{pr.reject.eq}) does not exceed a specified significance level, $\,\alpha$.
These parameters allow us to calculate the probability of early termination and the bound on the expected sample size given at~(\ref{ess.eq}).
The following section describes several such examples.

The power of the study is the probability of rejecting the null hypothesis.  
The probability in~(\ref{pr.reject.eq}) is a general statement, valid for any values of $\,(p_1, p_2)\,$ satisfying $\,0 \leq p_2\leq p_1\leq 1$.  
In addition to describing the statistical significance at the null hypothesis, we can consider other parameter values as well, and use this same expression to determine the power at other choices of $\,(p_1, p_2)$.
The designs proposed in the following section are geared towards demonstrating an increase in the value of the efficacy parameter $\,p_2\,$ while holding the safety parameter $\,p_1\,$ fixed.
Of course, we can also use~(\ref{pr.reject.eq}) to describe power against other alternative hypotheses as well, where $\,p_1\,$ as well as $\,p_2\,$ are allowed to vary between hypotheses.
The following section includes an example of varying both $\,p_1\,$ and $\,p_2$, under the alternative hypothesis.

We also have:
\begin{equation}                                       
   {\rm Power} \leq \Pr[\, {\rm Early \ stopping}\, ]
                                         \label{power.stop.eq}
\end{equation}
because we can not reject the null hypothesis if we decide to terminate the clinical trial early.
This bound proves to be critical in choosing a suitable design, as we show next.

\section{Design Guidance}
\label{guidance.section}

The previous section provides the mathematical background needed to assess the characteristics of any design and also allows us to compare one design with another.  
In this section we provide some guidance on how to select a useful and practical design.
We will demonstrate, in a specific example, that the same criteria used in selecting a Simon two-stage design is not always useful in the present setting.

Let us consider a specific set of parameter values in order to describe these guidelines.  
In this illustration we restrict designs to those with a maximum of $\,n_1+n_2=36\,$ patients to be enrolled.  
The figure of 36 is in line with the size of designs for Phase II studies that we regularly conduct at the Yale Comprehensive Cancer Center.  
Typical of Phase II studies, we also use a statistical significance of $\,\alpha=.1$.

To motivate the use of the methods developed in this work, consider the safety parameter $\,p_1=.8\,$ to model the PFS rate at follow-up time $\,t_1=2$~months of Stage~1.
We also assume $\,p_2=.2\,$ to describe the longer-term PFS rate at $\,t_2=6$~months in Stage~2.  
These times and rates are suggested by published survival curves~[\ref{Quoix 2005}, \ref{LaCaer 2012}] that we will use as the basis of our historical control values.  
All of these parameter values are typical for the type of studies that we conduct.

Let us begin by considering every possible choice of non-negative integer values of $\,n_1\,$ and $\,n_2\,$ such that $\,n_1+n_2=36$.  
For every choice of $\,n_1\,$ and $\,n_2\,$ we also look at all possible choices of cut-off values $\,r_1\,$ and $\,r_2\,$ and choose values that achieve one of several different characteristics that are detailed in this section.   
Table~\ref{design.detail.table} summarizes all of the designs selected and discussed in this section.  
In order to provide guidance on how to select a suitable design, let us begin by using those same criteria that are usually chosen when planning a Simon two-stage design~[\ref{Simon 1989}].

{\it Largest statistical significance}.  
We can not achieve a statistical significance of exactly $\,\alpha\,$ because of the discrete distributions involved. 
The exact significance level will always be slightly lower than this desired value.  
This results in a test that is considered `conservative' because the exact statistical significance is lower than the claimed $\,\alpha\,$ level.      
The design with the largest statistical significance, without exceeding $\,\alpha=.1,$ is labeled as `A' in Table~\ref{design.detail.table}. 
A test with lower statistical significance will typically be at a disadvantage when comparing its power with a test with a higher significance levels, all other things remaining equal.
This disadvantage, however, is only relevant to a small and unimportant range of the parameter values, as we see in Fig.~\ref{power.fig}, when we compare the power of this design with that of others that are suggested, below.

{\it Lowest expected sample size.}  
This design criteria is also called `optimal' in the language of traditional Simon two-stage Phase II designs.
This design is listed as `B' in Table~\ref{design.detail.table}.
Such optimal designs offer efficiency in terms of lower expected costs and numbers of patients accrued.
In Table~\ref{design.detail.table} we calculate the expected sample size using the upper bound provided at~(\ref{ess.eq}).

{\it Highest probability of early termination}.  
Also called `minimax' because this minimizes the probability of the largest possible sample size.  
This is listed as design `C' in Table~\ref{design.detail.table}.
These designs are similar to the optimal designs because these aim to minimize the maximum possible costs.

{\it Balanced}:  $\,n_1 = n_2$.  
In a balanced design, equal numbers of patients are enrolled in Stages~1 and~2.  
Intuitively, such a design places equal emphasis on safety and efficacy. 
This is listed as design `D' in Table~\ref{design.detail.table}.
Balanced designs suggest that we place equal effort into demonstrating both safety and efficacy.

In Table~\ref{design.detail.table} we can see that the four designs A--D all have a probability of early stopping greater than .89.  
If only $\,p_2\,$ varies between the null and alternative hypotheses, then the power can not exceed .11 in any of these designs because of the bound on power given by~(\ref{power.stop.eq}).
One way to raise the power is to consider other alternative hypotheses in which $\,p_1\,$ also increases.
This was not part of the original hypothesis being tested but could certainly be incorporated into the design for planning purposes.
In the following section we provide an example of how the power is affected by varying both $\,p_1\,$ and $\,p_2$.

Designs C and D in Table~\ref{design.detail.table} also have values of $\,r_2=0.$
This is an unrealistic parameter value because if we decide to progress from Stage~1 to Stage~2, then we will always reject the null hypothesis.
In other words, all of the statistical significance is determined in the safety portion of the trial, and not the efficacy portion.
This explains why the power of these designs is so low for testing an alternative hypothesis of increased value of $\,p_2$.
The designs A--D, based on the criteria that proved useful when planning a Simon two-stage trial all have serious shortcomings, in terms of low power.   
We will next introduce another set of useful criteria.

In light of~(\ref{power.stop.eq}), we propose limiting the probability of early stopping to values in the interval (.05, .2).
The designs proposed here provide separate measures of safety and efficacy. 
If the safety parameter of the proposed study is reasonably close to $\,p_1\,$ then we should not be too eager to terminate the study early.
Our principal concern in Stage~1 is testing whether the safety rate  $\,p_1\,$ is lower than the historical rate.  
In Fig.~\ref{pearly.fig} we plot the probability of early termination of the trial for different values of $\,p_1$.

The statistical significance can be relaxed and we should consider any value between (.085, .1) as reasonably close to acceptable.  
We want to make an early decision whether or not to progress from Stage~1 to Stage~2, so we also suggest that $\,n_1\,$ be much smaller than $\,n_2$.  
In the specific suggested designs (E--H) of Table~\ref{design.detail.table} we only include designs with $\,n_1\leq n_2/2$.     
For all designs with 36 patients, there are four designs that meet these suggested criteria and these are listed as E--H in Table~\ref{design.detail.table}.

Table~\ref{design.detail.table} illustrates the power for an alternative hypothesis of $\,p_2=.4\,$ for each of the designs listed.
Clearly, designs E--H have much greater power than A--D.
This difference is explained by the bound on power given at~(\ref{power.stop.eq}).   
The large difference in power is also demonstrated in Fig.~\ref{power.fig} where we can see the power at other values of $\,p_2$.  
There is a large difference between the power of designs A--D and E--H but negligible differences in power within each of these two groups of designs.

The probability of early stopping in Stage~1 is plotted in Fig.~\ref{pearly.fig} for the designs A--H listed in Table~\ref{design.detail.table}.
Designs A--D have high probabilities of stopping early at the historical PFS rate of $\,p_1=.8\,$ but this also results in greatly reduced power.
Designs E--H were chosen to have early stopping probabilities between .05 and .2 for a PFS rate of $\,p_1=.8\,$ at follow-up time $\,t_1=2$~months.
(These lmits are indicated with dotted lines.)  
Designs E--H have increasing numbers of observations $\,n_1\,$ assigned to Stage~1, and consequently have monotonically greater chances of stopping early at values of $\,p_1\,$ lower than .8.

Designs E--H have negligible differences in power so the choice of which one to use has more to do with our relative concerns for safety and the probability of inadvertently terminating the trial early for a safe drug.  Specifically, greater concerns for safety would prefer smaller values of $\,n_1\,$ so an early termination decision for a truly dangerous treatment would be reached sooner.

Our planning may also include varying both $\,p_1\,$ and $\,p_2\,$ between hypotheses.  
In this case, Fig.~\ref{powerF.fig} provides a comprehensive display of power for design F, as an example.
The null hypothesis in this case ($p_1=.8\,$ and $\,p_2=.2$) is indicated.  
In this figure, we can consider the power at any other pair of parameter values satisfying $\,0\leq p_2 \leq p_1 \leq 1$.

\section{Discussion}
\label{discussion.section}

To complete a discussion of designs for clinical trials, let us add three others that are not of the form described in  Section~\ref{notation.section}.  
These three designs (labeled X, Y, and Z) do not have separate criteria for stopping and efficacy: these two criteria are the same.  
In terms of the notation in Section~\ref{notation.section}, these three designs use PFS at $\,t_1=t_2\,$ and $\,p_1=p_2=.2$.  
The design `X' in Table~\ref{design.detail.table} is a simple binomial sampling with no early stopping and a sample of size 36 patients. If 11 or more patients have PFS at time $\,t_2\,$ (which occurs with probability .2 under the null hypothesis) then we reject that hypothesis at significance level .0889.

The last two designs listed in Table~\ref{design.detail.table} are traditional Simon, two-stage designs.  
Design `Y' is the ``optimal'' design with a maximum of 37 patients and design `Z' is the ``minimax'' design with 36 patients.  These two Simon designs and their powers were calculated using the {\bf clinfun} package in {\bf R}.

The Simon two-stage design~[\ref{Simon 1989}] is a special case of the designs proposed here with $\,t_1=t_2\,$ and $\,p_1=p_2$. 
The Simon design uses the same criteria for early stopping and efficacy.
Specifically, the traditional Simon design counts all of the safety successes ($X_1$) towards rejecting the null hypothesis, and, consequently should have greater power than the designs proposed here, but as we see in Fig.~\ref{power.fig}, this improvement is modest.
Similarly, the Simon designs with comparable sample sizes will also have greater probability of early termination under the null hypothesis, again, because early stopping and efficacy decisions are based on the same outcomes.
The binomial design (X) also has high power but lacks any early stopping rule.

The advantage of the designs proposed here (labeled E--H) allow the early termination decision to be made earlier and with a shorter delay between the accrual of Stages~1 and~2.  
The power of all designs in Table~\ref{design.detail.table} is plotted in Fig.~\ref{power.fig}.  
As pointed out earlier, designs A--D have negligible power.
In Fig.~\ref{power.fig} we see that the designs proposed (E--H) have slightly lower power than the binomial and Simon designs (X--Z) but as a trade-off, have a much shorter time to early stopping for safety.
Withing these groups of designs (E--H) and (X--Z), there are negligible differences in power.


\section*     {\bf References}

\begin{enumerate}


\item[\bibref{Simon 1989}]
Simon R.  Optimal two-stage designs for phase II clinical trials. {\it Controlled Clinical Trials\/} 1989; {\bf 10}: 1--10.

\item[\bibref{Thall 1990}]
Thall PF, Simon R.   Incorporating historical control data in planning phase II clinical trials. {\it Statistics in Medicine\/} 1990; {\bf 9}: 215--228.

\item[\bibref{Korn 2006}]
Korn EL, Freidlin B. Conditional power calculations for clinical trials with historical controls. {\it Statistics in Medicine\/} 2006; {\bf 25}: 2922--2931. DOI: 10.1002/sim.2516

\item[\bibref{Schill 2006}]
Schill W, Wild P.  Minmax designs for planning the second phase in a two-phase case–control study. {\it Statistics in Medicine\/} 2006; {\bf 25}: 1646--1659. DOI: 10.1002/sim.2307

\item[\bibref{Stallard 2012}]
Stallard N. Optimal sample sizes for phase II clinical
trials and pilot studies. {\it Statistics in Medicine\/} 2012; {\bf 31}: 1031--1042.

\item[\bibref{Hanfelt 1999}]
Hanfet JJ, Slack RS, Gehan EA.  A modification of Simon’s optimal design for phase II trials when the criterion is median
sample size. {\it Controlled Clinical Trials\/} 1999; {\bf 20}: 555--566.

\item[\bibref{DeWith 1983}]
De With C. Two-stage plans for the testing of binomial parameters. {\it Controlled Clinical Trials\/} 1983; {\bf 4}: 215--226.

\item[\bibref{Jung 2001}]
Jung SH, Carey M, Kim KM. Graphical search for two-stage designs for phase II clinical trials. {\it Controlled Clinical Trials\/} 2001; {\bf 22}: 367--372.



\item[\bibref{Gehan 1990}]
Gehan EA, Schneiderman MA. Historical and methodological developments in clinical trials at the National Cancer Institute. {\it Statistics in Medicine\/} 1990; {\bf 9}:  871--880.

\item[\bibref{Lachin 2005}]
Lachin JM. A review of methods for futility stopping based
on conditional power. {\it Statistics in Medicine\/} 2005; {\bf 24}: 2747--2764. DOI: 10.1002/sim.2151

\item[\bibref{Todd 2007}]
Todd S. A 25-year review of sequential methodology in clinical studies {\it Statistics in Medicine\/} 2007; {\bf 26}: 237--252. DOI: 10.1002/sim.2763

\item[\bibref{DeMets 2012}]
DeMets DL. Current development in clinical trials: Issues old and new. {\it Statistics in Medicine\/} 2012; {\bf 31}; 2944--2954. DOI: 10.1002/sim.5405.



\item[\bibref{Lee 2012}]
Lee JJ, Chu CT. Bayesian clinical trials in action. {\it Statistics in Medicine\/}  2012; {\bf 31}: 2955--2972.

\item[\bibref{Zhao 2009}]
Zhao L, Woodworth G. Bayesian decision sequential analysis with survival endpoint in phase II clinical trials.  {\it Statistics in Medicine\/} 2009; {\bf 28}: 1339--1352. DOI: 10.1002/sim.3544

\item[\bibref{Jung 2004}]
Jung S-H, Lee T, Kim KM, George SL. Admissible two-stage designs for phase II cancer clinical trials. {\it Statistics in  Medicine\/} 2004; {\bf 23}: 561--569.  DOI: 10.1002/sim.1600.

\item[\bibref{Westfall 1998}]
Westfall PH, Krishen A, Young SS. Using prior information to allocate significance levels for multiple endpoints. {\it Statistics in Medicine\/} 1998 {\bf 17}: 2107--2119. (1998)

\item[\bibref{Banerjee 2006}]
Banerjee A, Tsiatis AA. Adaptive two-stage designs in phase II clinical trials. {\it Statistics in Medicine\/} 2006; {\bf 25}: 3382--3395. DOI: 10.1002/sim.2501



\item[\bibref{Kunz 2012}]
Kunz CU, Kieser M.  Estimation of secondary endpoints in two-stage phase II oncology trials  {\it Statistics in Medicine\/}  2012; {\bf 31}: 4352--4368. DOI: 10.1002/sim.5585

\item[\bibref{Bowden 2012}]
Bowden J,  Wason J. Identifying combined design and analysis procedures in two-stage trials with a binary end point. {\it Statistics in Medicine\/} 2012; {\bf 31}; 3874--3884. DOI: 10.1002/sim.5468



\item[\bibref{Ensign 1994}]
Ensign, LG, Gehan EA, Kamen DS, Thall PF.  An optimal three-stage design for phase II clinical trials. {\it Statistics in Medicine\/} 1994; {\bf 13}: 1727--1736.

\item[\bibref{Zhong 2012}]
Zhong W, Koopmeiners JS, Carlin BP.  A trivariate continual reassessment method for phase I/II trials of toxicity, efficacy, and surrogate efficacy. {\it Statistics in Medicine\/} 2012; {\bf 31}; 3885--3895. DOI: 10.1002/sim.5477



\item[\bibref{Hong 2012}]
Hong S, Shi L. Predictive power to assist phase 3 go/no
go decision based on phase 2 data on a different endpoint. {\it Statistics in Medicine\/} 2012; {\bf 31}: 831--843

\item[\bibref{Chang 2007}]
Chang MN, Devidas M, Anderson J. One- and two-stage designs for phase II window studies.  {\it Statistics in Medicine\/} 2007; {\bf 26}: 2604--2614. DOI: 10.1002/sim.2741



\item[\bibref{Thall 1999}]
Thall PF, Cheng S-C. Treatment comparisons based on two-dimensional safety and efficacy alternatives in oncology trials. {\it Biometrics\/} 1999; {\bf 55}: 746--753

\item[\bibref{Thall 2001}]
Thall PF, Cheng S-C. Optimal two-stage designs for clinical trials based on safety and efficacy.  {\it Statistics in Medicine\/} 2001; {\bf 20}: 1023--1032. DOI: 10.1002/sim.717

\item[\bibref{Shih 2004}]
Shih WJ, Ohman-Strickland PA, Lin Y. Analysis of pilot and early phase studies with small sample sizes. {\it Statistics in Medicine\/} 2004; {\bf 23}: 1827--1842. DOI: 10.1002/sim.1807



\item[\bibref{Quoix 2005}]
Quoix E, Breton J-L, Ducolon´e A, Mennecier B, Depierre A, Lemari´e E, Moro-Sibilot D, Germa C,  Neidhardt A-C. First line chemotherapy with gemcitabine in advanced non-small cell lung cancer elderly patients: A randomized phase II study of 3-week versus 4-week schedule. {\it Lung Cancer\/} 2005; {\bf 47}: 405--412.

\item[\bibref{LaCaer 2012}]
LeCaer H, Greillier  L, Corre R, Jullian H, Crequit J, L. Falcherof, C. Dujong, H. Berardh, A. Vergnenegrei, C. Chouaidj, the GFPC 0505 Team1. A multicenter phase II randomized trial of gemcitabine followed by erlotinib at progression, versus the reverse sequence, in vulnerable elderly patients with advanced non small-cell lung cancer selected with a comprehensive geriatric assessment (the GFPC 0505 study). {\it Lung Cancer\/} 2012; {\bf 77}: 97--103.



\item[\bibref{Uppuluri 1970}] 
Uppuluri VRR and Blot WJ (1970). ``A probability distribution arising in a riff-shuffle.'' {\it Random Counts  in Scientific Work, 1: Random Counts in Models and Structures},  G.P. Patil (editor), University Park: Pennsylvania State University Press, pp  23--46.

\item[\bibref{Zhang 2000}]
Zhang Z, Burtness BA and Zelterman D.  The maximum negative binomial distribution. {\em Journal of Statistical Planning and Inference\/}  2000; {\bf 87}: 1--19.

\item[\bibref{Zelterman 2004}]
D Zelterman (2004). {\it Discrete Distributions: Application in 
the Health Sciences}, New York: J. Wiley. xix + 277pp

\end{enumerate}


\section*{Appendix: Distribution of the Smallest Sample Size in Stage~1}

We may be able to decide early either to continue to Stage~2 or else to terminate before all of the $\,n_1\,$ Stage~1 patients are enrolled.
Specifically, as soon as either $\,r_1\,$ Stage~1 successes or else $\,n_1-r_1+1\,$ failures are observed then we have sufficient data to make the decision whether to continue or terminate (respectively) without having to wait to observe the remainder of the Stage~1 subjects.

Begin by considering a more general result.  
Let $\,Z_1, Z_2, \ldots \,$ denote  a sequence of independent, identically distributed Bernoulli random variables with $\,\Pr[Z_i =1 ]=p\,$ for probability parameter $\,p\; (0\leq p\leq 1)$. 
For non-negative integer values $\,s\,$ and $\,t,$ let $\,Y\,$ denote the smallest integer value such that $\,\{Z_1, \ldots , Z_Y\}\,$ contains either $\,s\,$ ones (successes) or else $\,t\,$ zeros (failures).

The distribution of $\,Y\,$ has support on integer values in the range 
\begin{equation}                                     
                \min(s,t) \leq \; Y \;\leq s+t-1 . \label{range.y.eq}
\end{equation}

If $\,Z_Y=1\,$ then this represents the $s-$th success and $\,Y-s\,$ has a negative binomial distribution with
\begin{equation}                                    
   \Pr[\,Y=s+j {\rm \ \ and \ \ } Z_Y=1\,]        \label{nb1.eq}
          = {{s+j-1}\choose{s-1}} p^s (1-p)^j
\end{equation}
for $\,j=0, 1, \ldots$.

Similarly, if $\,Z_Y=0\,$ then $\,Y-t\,$ has a negative binomial distribution with
\begin{equation}                                    
    \Pr[\,Y=t+j {\rm \ \ and \ \ } Z_Y=0\,]        \label{nb2.eq} 
               = {{t+j-1}\choose{t-1}} p^j(1-p)^t 
\end{equation}
for $\,j=0, 1, \ldots$.

The expressions~(\ref{nb1.eq}) and~(\ref{nb2.eq}) represent probabilities of mutually exclusive events so the distribution of $\,Y\,$ is found as the sum of these two probabilities, restricted to the range of $\,Y\,$ satisfying~(\ref{range.y.eq}).  
This derivation was verified by extensive simulations, not presented here. 
For the special case of $\,s=t,$ the distribution of $\,Y\,$ is the riff-shuffle, or minimum negative binomial distribution~[\ref{Uppuluri 1970}].
Similar derivations of the closely related maximum negative binomial discrete distributions also appear in~[\ref{Zhang 2000}, \ref{Zelterman 2004}].
The distribution of $\,Y\,$ does not have a simple, closed form but can easily be calculated numerically.

From this general result, we can find an expression for the expectation of the minimum number of patients necessary in Stage~1 in order to reach any early decision. 
We obtain the distribution of the number of these patients necessary to either continue on to Stage~2 or else terminate the study early by setting $\,p=p_1,\;s=r_1\,$ and $\,t=n_1-r_1+1$.

Table~\ref{Exp.Stage1.table} gives the expected value and standard deviation of the minimum number of patients required to reach either decision in Stage~1 for designs E--H.


\begin{table}[p]
\caption{Design parameters, outcomes, their ranges, and interpretations 
     \label{parameters.table}}
\begin{center}
\begin{tabular}{c l l} \\[-5ex]\hline
Parameters & Ranges & Interpretation\\ \hline
$t_1,\; t_2$   & $0 \leq t_1 \leq t_2$ & Follow-up times on study in Stage~1 and Stage~2 \\[1ex]
                
$n_1, \; n_2$   & Non-negative & Sample sizes in Stages 1 and 2 \\
                      & integers \\[1ex]
$p_1, \; p_2$ &$0\leq p_2\leq p_1\leq 1$ 
                      & Probability of successful outcomes\\
                  &&  at follow-up times $\,t_1\,$ and $\,t_2$. \\[1ex]
$r_1,\; r_2$ & Non-negative 
           & $r_1\,$ is the minimum number of successes in at time $\,t_1$  \\
      & integers & \ \ \ \ in Stage~1 for continuing from Stage~1 to Stage~2 \\[1ex]
        & $0 \leq r_1 \leq n_1$ & $r_2\,$ is the minimum number of 
                 successes at  follow-\\
    & $0\leq r_2 \leq n_1+n_2$ & \ \ \ \ up time $\,t_2\,$ 
           needed to reject the null hypothesis \\  \hline
Outcome \\ variables \\ \hline
$X_i$ & $0,1,\ldots, n_i$ 
          & Number of the successes in Stage $\,i$\\
    & \quad$i=1,2$ & Independently distributed as Binomial $(n_i, \,p_i)$ \\[1ex]
$X_{12}$ & $0,1,\ldots, X_1$ & Number of the $\,X_1\,$ 
                     successes at time $\,t_1\,$ in Stage~1 who \\
    && later become Stage~2 successes at follow-up time $\,t_2$. \\
   && Distributed as Binomial 
                 $(X_1,\, p_2/p_1)$ given $\,X_1$ \\ \hline
\end{tabular}
\end{center}
\end{table}

\newpage
\begin{table} [p]                          
\caption{Details of selected designs with no more than $\,n_1+n_2= 36\,$ patients, maximum significance level of .1, safety parameter $\,p_1=.8,$ and null hypothesis $\,p_2=.2$.  
Designs A--D follow traditional guidelines and designs E--H are suggested  in Section~\ref{guidance.section}.  
The expected sample size is the upper bound given at~(\ref{ess.eq}).
The power is for the alternative hypothesis with $\,p_2=.4$.  \label{design.detail.table}}
\begin{center}
\begin{tabular}{l cccc cccc} \hline
&&&&&& Expected & Probability & Power\\
Label:&&&&& Exact & sample & of early & for \\
Description & $n_1$ & $n_2$ & $r_1$ & $r_2$ & $\alpha$ 
        & size & stopping & $p_2=.4$\\ \hline

A: Highest $\,\alpha\,$ & 
31 & \p5 & 28 & \p5 & \p.09997 & 31.54 & .893 & .107 \\
B: Lowest expected  \\ sample size  &
\p9 & 27 & \p9 & \p7 & .0898 & 12.62 & .866 & .134\\
C: Highest probability \\of early termination &
 32 & \p4 &29 & \p0 & .0931 & 32.37 & .907 & .093 \\
D: Balanced: $n_1=n_2$ &
18 &18& 17 & \p0 & .0991 & 19.78 & .901 & .099 \\ \hline

\multicolumn{4}{l}{Suggested designs:} \\ \hline
E: &\p5 & 31 & 3 & 11 & .0858 & 34.20 & .058 & .861\\
F: &\p8 & 28 & 5 & 11 & .0862 & 34.42 & .056 & .863\\
G: & 11 & 25 & 7 & 11 & .0868 & 34.74 & .050 & .869\\
H: & 12 & 24 & 8 & 11 & .0858 & 34.26 & .073 & .851 \\ \hline
\multicolumn{9}{l}{Other designs for $\,t_1=t_2\,$ and $\,p_1=p_2=.2$.}
  \\ \hline
X: Binomial &  --  & 36 & -- & 11 & .0889 & 36.\q & 0 & .910\\
Y: Simon optimal & 17 & 10 & 3 & 10 & .0948 & 26.02 & .549 & .903\\
Z: Simon minimax & 19 & 15 & 3 & 10 & .0861 & 28.26 & .455 & .902 \\ \hline
\end{tabular}
\end{center}
\end{table}


\begin{table} [p!]   
\vspace*{2in}
\caption{Expected value and standard deviation of the minimum number of patients necessary to reach any decision in Stage~1 with $\,p_1=.8$.  The distribution is derived in the Appendix. \label{Exp.Stage1.table}}
\begin{center}
\begin{tabular}{c cc cc} \\ \hline
 &&& Expected & Standard \\
Design & $n_1$ & $r_1$ &  value & deviation \\ \hline
E & \p5  &  3 & 3.63 & \p.73  \\
F & \p8  &  5 & 6.11 & 1.00 \\
G &  11  &  7 & 8.60 & 1.23 \\
H &  12  &  8 & 9.76 & 1.26 \\ \hline
\end{tabular}
\vspace*{2in}
\end{center}
\end{table}



\newpage              

\typeout{}\typeout{Figure 1: Graphical schema of notation}\typeout{}
\begin{figure}[p!!]

\beginpicture
\setcoordinatesystem units <1in, 1in> \setplotarea x from 0 to 6, y from 0 to 3.5 \put { \ } at -.25 1

\setlinear 
\thinlines 
\put {Time on study:}[l]  at 0 3.4
\put {\shortstack{Stage~1: \\
              $n_1\,$ patients}} at .5 2.5
\put {\shortstack{Stage~2: \\
              $n_2\,$ patients}} at .5 1
\plot 0 3.2 5.5 3.2 /  

\putrectangle corners at 0 2.85 and 1 2.15 
\putrectangle corners at 0 1.35 and 1 0.65

\put{$1-p_1$} at 1.75 2.75
\put{$p_1$} at 1.75 2.1


\put {$t_1$}[l] at 3 3.4
\put {\shortstack{$n_1-x_1\,$ \\failures}} at 3 2.75

\put {\shortstack{$x_1$ \\ successes}} at 3 2
\plot 2.5  2.75  1 2.5 2.5 2 /


\put {$t_2$} at 5   3.4

\put{\shortstack{$x_1-x_{12}$ \\ failures}} at 5 2.25
\put{\shortstack{$x_{12}$ \\ successes}} at 5 1.75
\put{\shortstack{$x_2$ \\ successes}} at 5 1.25
\put{\shortstack{$n_2-x_2$ \\ failures}} at 5 0.75
\put{$p_2$} at 3 1.25
\put{$1-p_2$} at 3 0.75
\put{$1-p_2/p_1$} at 4 2.35
\put{$p_2/p_1$} at 4 1.75

\plot 4.5 2.25 3.5 2 4.5 1.75 /
\plot 4.5 1.25 1 1 4.5 0.75 /
\plot 5.25 1.9  5.4 1.9 5.4 1.1 5.25 1.1 /
\plot 0 0.25 5.5 0.25 / 
\endpicture

\caption{Graphical schematic of notation for follow-up times, numbers of patients enrolled, outcome classifications and their binomial probabilities.  
At the conclusion of Stage~1, at follow-up time $\,t_1$, there are $\,x_1\,$ successes and we need a minimum of $\,r_1\,$ of these to continue to Stage~2.  
At follow-up time $\,t_2\,$ there are $\,x_{12}+x_2\,$ successes and we need a minimum of $\,r_2\,$ of these to reject the null hypothesis. \label{schema.fig}}

\end{figure}


\newpage              
\typeout{}\typeout{Figure 2: All designs}\typeout{}
\begin{figure}[p!!]
\begin{center}
\includegraphics[scale=.9]{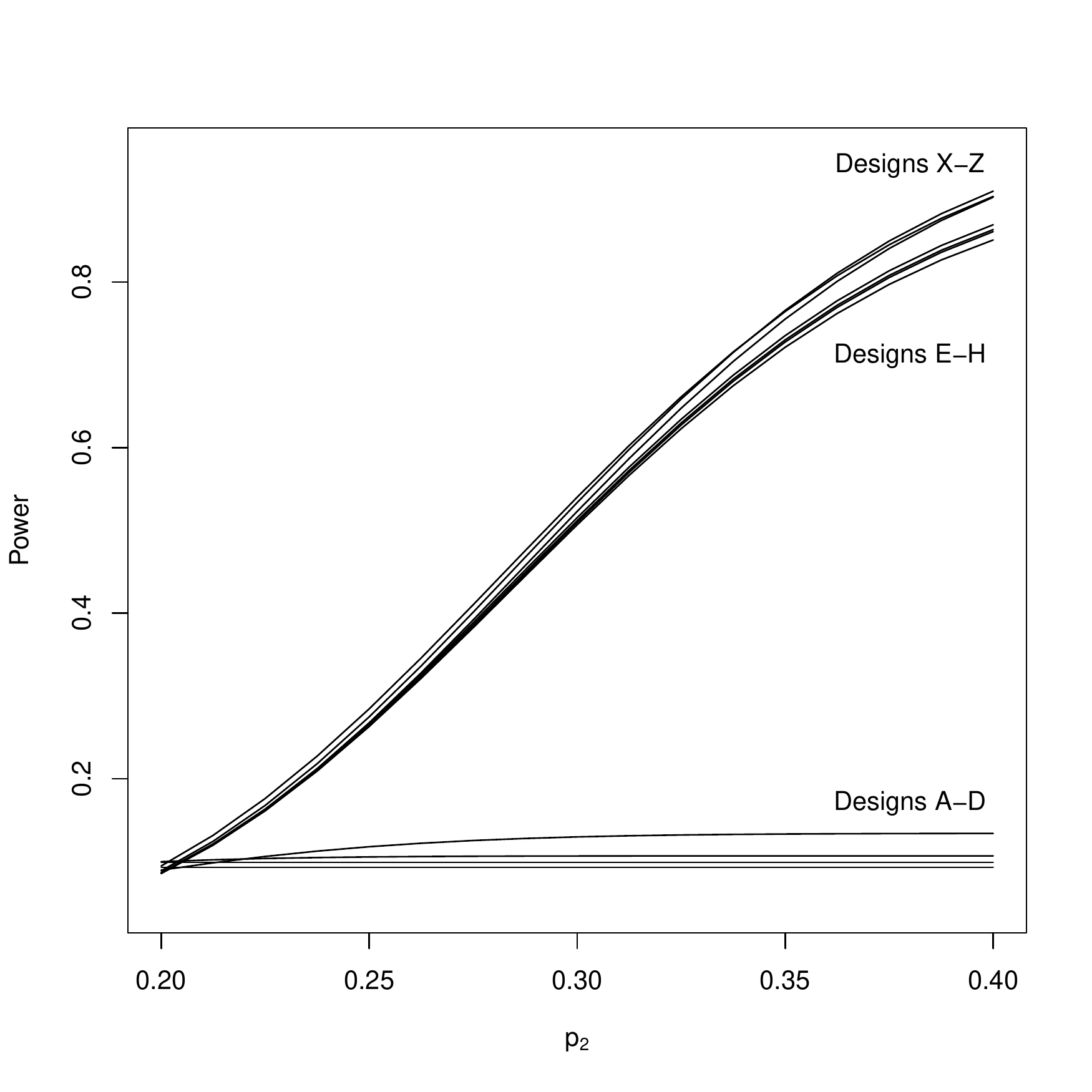}
\end{center}
\caption{Power of designs listed in Table~\ref{design.detail.table} for a fixed value of the safety parameter $\,p_1=.8$.  The null hypothesis is $\, p_2=.2\,$ and the significance level is $\,\alpha=.1$.
\label{power.fig}}
\end{figure}


\newpage        
\typeout{}\typeout{Figure 3: Prob of early termination}\typeout{}
\begin{figure}[p!!]
\begin{center}
\includegraphics[scale=.9]{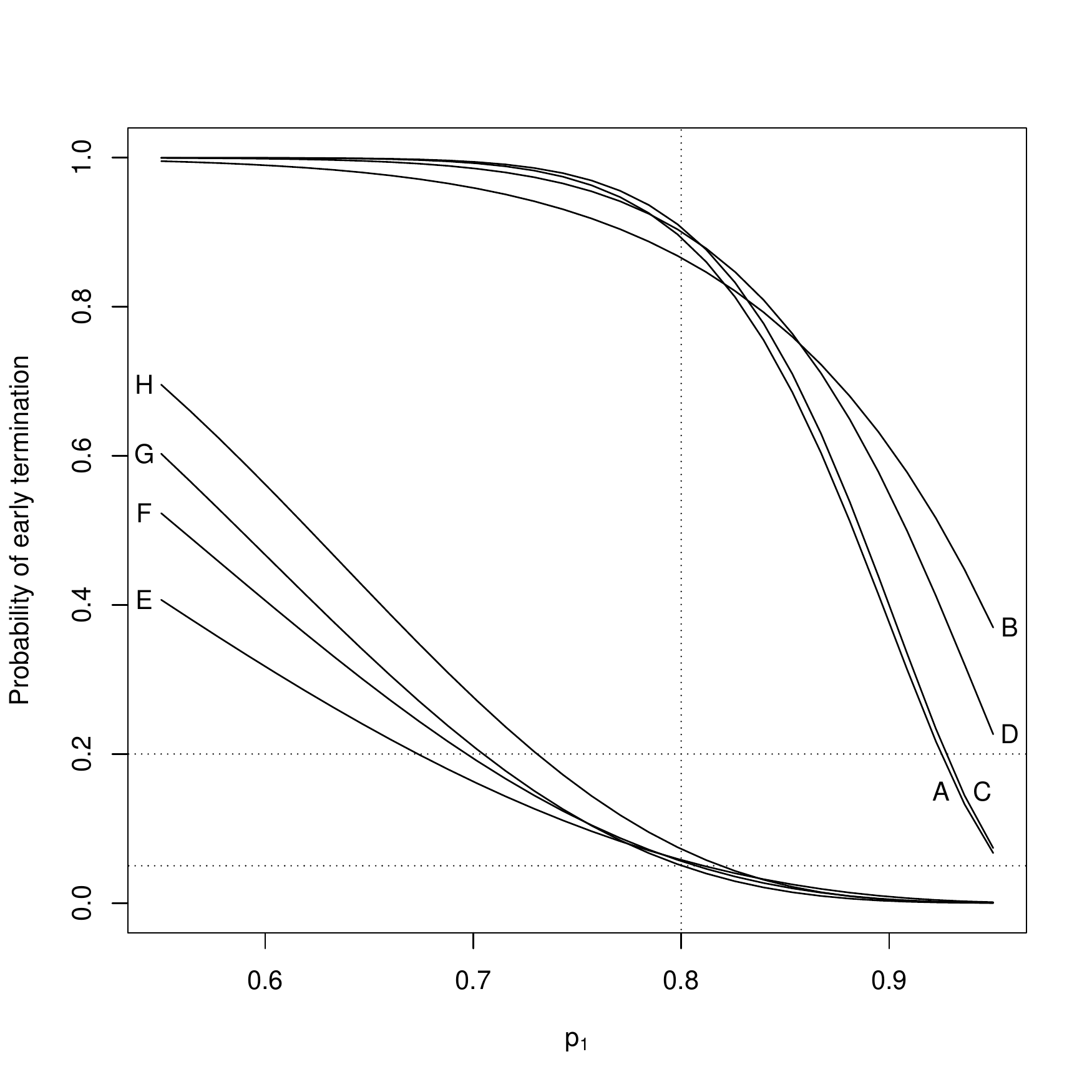}
\end{center}
\caption{Probability of early termination for designs in Table~\ref{design.detail.table}. 
Suggested designs E--H are selected to have early termination probabilities between .05 and .2 at $\,p_1=.8$.  
These limits are given with dotted lines.
\label{pearly.fig}}
\end{figure}


\newpage              
\typeout{}\typeout{Figure 3: Design F}\typeout{}
\begin{figure}[p!!]
\begin{center}
\includegraphics[scale=.9]{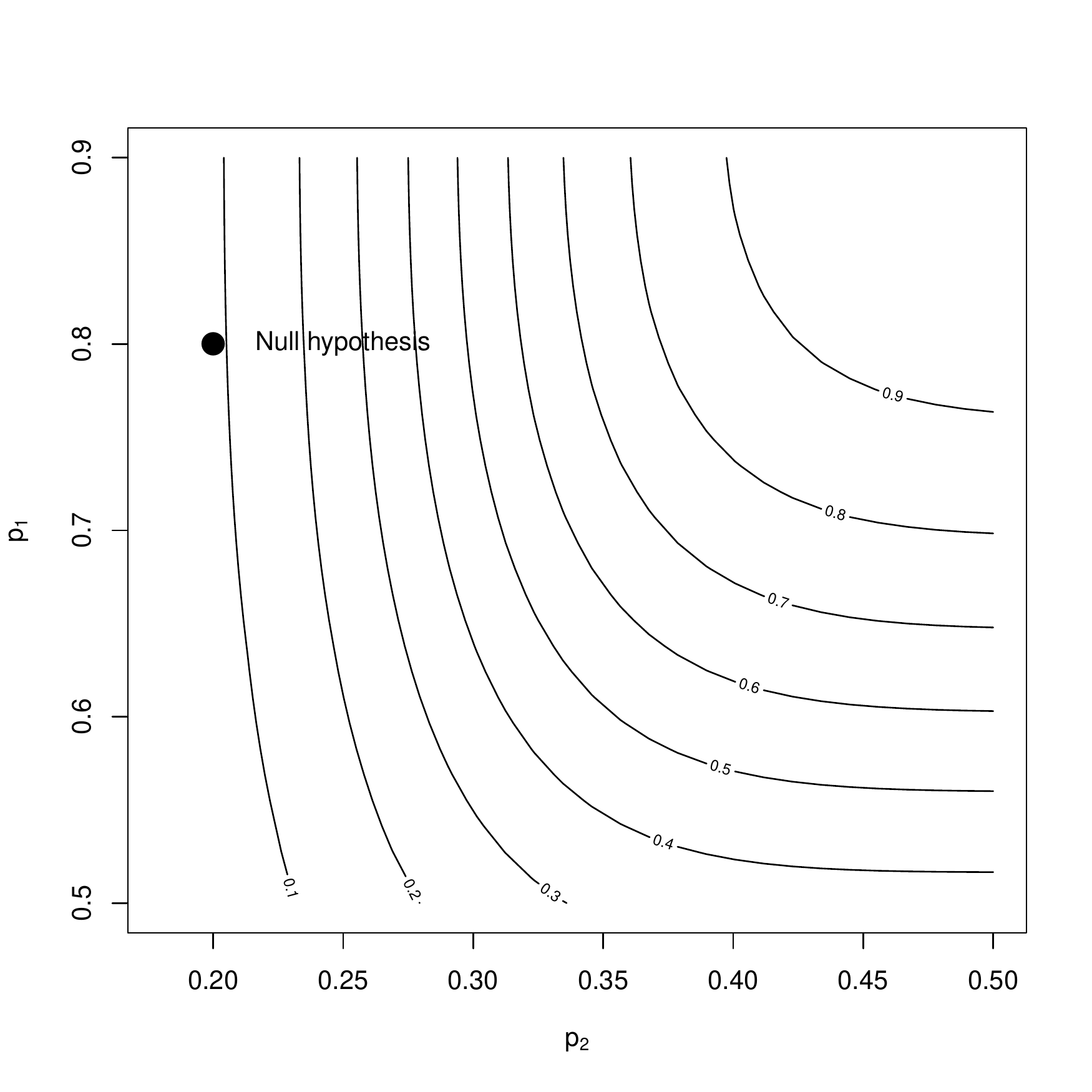}
\end{center}
\caption{Probability of rejecting the null hypothesis for design `F' in Table~\ref{design.detail.table}.  The location of the null hypothesis is indicated.  \label{powerF.fig}}
\end{figure}


\end{document}